\documentclass[fleqn,usenatbib]{mnras}

\usepackage[T1]{fontenc}
\usepackage{ae,aecompl}

\usepackage{epsf,palatino,url,graphicx,wrapfig,array,setspace,amsmath,amssymb,fancyhdr,multirow,lscape,appendix,rotating}
\usepackage{graphics,epsfig,txfonts}

\usepackage{longtable}
\usepackage{amssymb}
\usepackage{xcolor}
\usepackage{color}
\usepackage{lipsum}
\usepackage{textcomp}
\usepackage{threeparttable}
\usepackage{enumerate}
\usepackage{placeins}
\usepackage{float}
\usepackage[normalem]{ulem}
\usepackage{soul}
\usepackage{blindtext}
\usepackage[hyphenbreaks]{breakurl}

\usepackage{tikz,hyperref}
\definecolor{lime}{HTML}{A6CE39}
\DeclareRobustCommand{\orcidicon}{%
	\begin{tikzpicture}
	\draw[lime, fill=lime] (0,0) 
	circle [radius=0.16] 
	node[white] {{\fontfamily{qag}\selectfont \tiny ID}};
	\draw[white, fill=white] (-0.0625,0.095) 
	circle [radius=0.007];
	\end{tikzpicture}
	\hspace{-2mm}
}

\foreach \x in {A, ..., Z}{%
	\expandafter\xdef\csname orcid\x\endcsname{\noexpand\href{https://orcid.org/\csname orcidauthor\x\endcsname}{\noexpand\orcidicon}}
}

\newcommand{\orcid}[1]{\href{https://orcid.org/#1}{\textcolor[HTML]{A6CE39}{\orcidicon}}}

\newcommand{\nicer}{\textit{NICER}\xspace}

\newcommand{\glast}{{\it Fermi}\xspace}
\newcommand{\fermi}{{\it Fermi}\xspace}
\newcommand{\swift}{{\it Swift}\xspace}

\newcommand{\nustar}{{\it NuSTAR}\xspace}
\newcommand{\nus}{{\it NuSTAR}\xspace}

\newcommand{\ergs}[1]{$\times 10^{#1}$ erg s$^{-1}$}
\newcommand{\oergs}[1]{$10^{#1}$ erg s$^{-1}$}

\newcommand{\sxp}{SXP\,9.3\xspace}
\newcommand{\rxj}{RX\,J0209.6$-$7427\xspace}

\newcommand{\eqb}{\begin{eqnarray}}
\newcommand{\eqe}{\end{eqnarray}}

\title[2019 outburst of RX\,J0209.6$-$7427]{The 2019 super-Eddington outburst of RX\,J0209.6$-$7427: Detection of pulsations and constraints on the magnetic field strength}

\author[G. Vasilopoulos et al.]
{G.~Vasilopoulos\orcid{0000-0003-3902-3915},$^1$\thanks{E-mail: georgios.vasilopoulos@yale.edu}
P.~S.~Ray\orcid{0000-0002-5297-5278},$^{2}$
K.~C.~Gendreau,$^{3}$
P.~A.~Jenke,$^{4}$
G.~K.~Jaisawal\orcid{0000-0002-6789-2723},$^{5}$
\newauthor
C.~A.~Wilson-Hodge\orcid{0000-0002-8585-0084},$^{6}$
T.~E.~Strohmayer\orcid{0000-0001-7681-5845},$^{7}$
D. Altamirano\orcid{0000-0002-3422-0074},$^{8}$
W.~B.~Iwakiri,$^{9}$
\newauthor
M.~T.~Wolff\orcid{0000-0002-4013-5650},$^{2}$
S.~Guillot\orcid{0000-0002-6449-106X},$^{10}$
C.~Malacaria\orcid{0000-0002-0380-0041},$^{6,11}$\thanks{NASA Postdoctoral Fellow}
A.~L.~Stevens$^{12,13}$
\\
$^1$Department of Astronomy, Yale University, PO Box 208101, New Haven, CT 06520-8101, USA \\
$^2$Space Science Division, U.S. Naval Research Laboratory, Washington, DC 20375, USA\\
$^3$X-Ray Astrophysics Laboratory, NASA Goddard Space Flight Center, Greenbelt, MD 20771, USA\\
$^4$University of Alabama in Huntsville, Huntsville, AL 35805, USA\\
$^{5}$National Space Institute, Technical University of Denmark, Elektrovej 327-328, DK-2800 Lyngby, Denmark\\
$^6$Astrophysics Branch, NASA Marshall Space Flight Center, Huntsville, AL 35812, USA\\
$^7$Astrophysics Science Division and Joint Space-Science Institute,
NASA's Goddard Space Flight Center, Greenbelt, MD 20771, USA\\
$^{8}$School of Physics and Astronomy, University of Southampton, Southampton, SO17 1BJ, UK\\
$^9$Department of Physics, Faculty of Science and Engineering, Chuo University, 1-13-27 Kasuga, Bunkyo-ku, Tokyo 112-8551, Japan \\
$^{10}$IRAP, CNRS, 9 avenue du Colonel Roche, BP 44346, F-31028 Toulouse Cedex 4, France\\
$^{11}$Universities Space Research Association, NSSTC, 320 Sparkman Drive, Huntsville, AL 35805, USA\\
$^{12}$Department of Physics \& Astronomy, Michigan State University, 567 Wilson Road, East Lansing, MI 48824, USA\\
$^{13}$Department of Astronomy, University of Michigan, 1085 South University Avenue, Ann Arbor, MI 48109, USA\\}

\date{ }

\pubyear{2020}

\begin{document}
\label{firstpage}
\pagerange{\pageref{firstpage}--\pageref{lastpage}}
\maketitle

\begin{abstract}
In November 2019, MAXI detected an X-ray outburst from the known Be X-ray binary system \rxj located in the outer wing of the Small Magellanic Cloud. We followed the outburst of the system with \nicer which led to the discovery of X-ray pulsations with a period of 9.3~s. We analyzed simultaneous X-ray data obtained with \nustar and \nicer allowing us to characterize the spectrum and provide an accurate estimate of its bolometric luminosity. During the outburst the maximum broadband X-ray luminosity of the system reached 1--2\ergs{39}, thus exceeding by about one order of magnitude the Eddington limit for a typical 1.4 $M_{\odot}$ mass neutron star (NS). Monitoring observations with \textit{Fermi}/GBM and \nicer allowed us to study the spin evolution of the NS and compare it with standard accretion torque models. We found that the NS magnetic field should be of the order of 3$\times10^{12}$ G. We conclude that \rxj exhibited one of the brightest outbursts observed from a Be X-ray binary pulsar in the Magellanic Clouds, reaching similar luminosity level to the 2016 outburst of SMC\,X-3. 
Despite the super-Eddington luminosity of \rxj, the NS appears to have only a moderate magnetic field strength.
\end{abstract}




\section{Introduction}

High mass X-ray binaries (HMXBs) are young binary systems where the massive companion (typically $>8M_{\odot}$) transfers material onto a compact object.
A major subclass of HMXBs are Be X-ray binaries (BeXRBs), a population that hosts the majority of the known X-ray pulsars \citep[][for a review on BeXRBs]{2011Ap&SS.332....1R}.
Among BeXRBs the vast majority host a Neutron Star (NS) as a compact object. At the moment  MWC\,656 is the only black-hole BeXRB known \citep[][]{2014Natur.505..378C}, while a couple candidate White Dwarf BeXRBs have been reported \citep[e.g.][]{1995A&A...296..685H,2001A&A...377..148T,2012A&A...537A..76S}.
In BeXRBs the donor is a Be star that loses material through a slow moving equatorial wind that is referred to as decretion disk. As the binary plane can be misaligned to the decretion disk or the NS orbit can be highly eccentric, mass transfer is not constant and thus BeXRBs are typically highly variable systems. 
X-ray outbursts are typically observed as the NS passes through the Be disk.
Type I or normal outbursts have duration shorter than the orbital period and reach luminosity $\sim$10\% the Eddington limit for a NS ($L_{\rm Edd}=2\times10^{38}$ erg/s, assuming a NS mass of 1.4 $M_\odot$). Type II or giant outbursts are less frequent and can last several orbits while reaching luminosities in excess of $L_{\rm Edd}$ \citep[see ][for various outbursts mechanisms]{2001A&A...377..161O,2013PASJ...65...41O,2014ApJ...790L..34M}.

In X-ray pulsars, the strong magnetic field of the NS disrupts the flow of matter at a distance where the magnetic field pressure equals the ram pressure of the flow.
Material is then funneled along the field  lines  onto  the  magnetic pole, forming the so-called accretion column \citep{1976MNRAS.175..395B}. The emission characteristics of this region depend on the mass accretion rate, as for large values a shock develops above the NS surface.
By following the evolution of X-ray outbursts in BeXRBs we can gain insight on the radiative processes in the accretion column and the formation of the shock  above the NS surface that is believed to define the transition between subcritical and supercritical accretion regimes.
For many systems, this transition has been observationally determined to occur close to X-ray luminosities of \oergs{37} \citep[see characteristic example of EXO 2030,][]{2013A&A...551A...1R,2017MNRAS.472.3455E}, as below (above) this critical $L_X$ the spectra of the systems become harder (softer) when brighter (typically within 2.0--10.0 keV band).

In addition, major outbursts of BeXRBs have historically offered the first evidence that accretion onto NSs can at least momentarily exceed $L_{\rm Edd}$. However, the discovery of pulsations from M82~X-2, a system with luminosity of $100\times{L}_{\rm Edd}$, demonstrated that stable accretion onto NSs at super-Eddington rate is possible \citep[][]{2014Natur.514..202B}. 
This discovery introduced a new category of systems, the so-called pulsating ultra-luminous X-ray sources (PULXs), that are broadly defined as extragalactic accreting NSs with luminosities greater than \oergs{39} \citep[see ][for a review on ULXs]{2017ARA&A..55..303K}. 
This realization has fueled a search that led to the discovery and study of more PULXs in recent years
\citep[e.g.,][]{2016ApJ...831L..14F,2017Sci...355..817I,2018MNRAS.476L..45C,2019arXiv190604791R,2019MNRAS.488L..35S}.
Furthermore, based on spectral similarities between non-pulsating and pulsating ULXs, there is now compelling evidence that a significant fraction of ULXs may host highly magnetized NSs \citep{2017A&A...608A..47K,2017ApJ...836..113P,2018ApJ...856..128W}.
In the broad sense, the brightest giant outbursts from BeXRBs ($L_{\rm X}>$\oergs{39}) temporarily qualify these systems as PULXs. However, there are a few key differences between the two categories of systems. Although both systems can show large variability in their X-ray flux (factor $>$100), in BeXRBs this is a result of variable mass transfer, while in extragalactic PULXs, mass transfer is stable through many orbits and variations in their observed flux occur in quasi-periodic super-orbital time scales, and are thought to be related to obscuration due to disk precession \citep{2017MNRAS.466.2236D,2017ApJ...834...77F,2019MNRAS.488.5225V,2019MNRAS.489..282M}.

In order to explain the super-Eddington luminosities of PULXs, it has been speculated \citep[][]{2015MNRAS.454.2539M} that the NSs in these systems must have high magnetic field strengths ($B>10^{13}$ G), which is at least an order of magnitude larger than NSs in typical X-ray pulsars \citep{2014MNRAS.437.3664H}. 
Albeit, in the recent work of \citet{2019MNRAS.485.3588K}, the authors claim that the observed properties of many PULXs can be explained by NS with magnetic field strength between $10^{11}$--$10^{13}$G.
A characteristic example is the PULX system NGC300\,ULX1 where the NS magnetic field (dipole term of $\sim10^{12}$ G) has been constrained by both timing studies \citep{2018A&A...620L..12V,2019MNRAS.488.5225V} and the possible detection \citep[see caveats,][]{2019A&A...621A.118K} of a cyclotron line \citep{2018ApJ...857L...3W}.
However, there are at least a couple of systems that their temporal properties probe magnetic fields similar to magnetar values \citep[e.g., M81\,X-2 and M51\,ULX-7,][]{2020ApJ...891...44B,2020MNRAS.491.4949V}.
Nevertheless, the radiative mechanisms of the NS accretion column \citep{2007ApJ...654..435B} have not yet been fully studied at super-Eddington accretion rates, where several assumptions break due to the accretion column geometry \citep{2017ApJ...835..129W}.

Given that the handful of persistent PULXs that are known lie at distances of a few Mpc, their detailed study is hampered by limitations of current X-ray observatories. 
While this might change with the launch of future proposed missions \citep[e.g., STROBE-X and eXTP,][]{2019arXiv190303035R,2019SCPMA..6229502Z}, at the moment the best laboratories to study super-Eddington accretion are major outbursts of BeXRBs in our local galaxy group. 
Perhaps the brightest X-ray outburst of a BeXRB was the 2017 outburst of the Galactic pulsar Swift\,J0243.6+6124 making it the first Galactic PULX \citep[luminosity of $\sim$2\ergs{39};][]{2018ApJ...863....9W}. 
However, an ideal place to study BeXRBs are the nearby star-forming Magellanic Clouds (MCs) galaxies. Given the known distance of the MCs and low foreground Galactic absorption in their direction, outbursts of MC pulsars  \citep[e.g., see SMC X-2, SMC X-3, LXP 8.04;][]{2016MNRAS.458L..74L,2017A&A...605A..39T,2018A&A...614A..23K,2014A&A...567A.129V}
can offer unique insight into super-critical accretion; i.e. the critical $L_{\rm X}$ where the shock is created above the NS and the accretion column is formed.

\rxj is a BeXRB system discovered by analysis of archival ROSAT PSPC observations \citep{2005A&A...435....9K}, and located in the outer wing of the Small Magellanic Cloud (SMC).
The only two historic outbursts occurred in March and November 1993 and were detected by ROSAT PSPC \citep{2005A&A...435....9K}. 
Both outbursts reached a luminosity of $\sim$\oergs{38} (0.1--2.4 keV band) and lasted approximately one month, while the outburst peaks were separated by about 200 days. 
In 2019, \rxj exhibited a major outburst where the X-ray luminosity of the system exceeded \oergs{39}, thus the system temporarily became a PULX. 
Given the proximity of the system, and the accurately measured distance of the SMC, this offers an ideal opportunity to study its properties and compare them with PULXs and other HMXB pulsars that have exhibited super-Eddington outbursts.

In this paper we present the first results of the X-ray monitoring of the system during its 2019 outburst (see Section~\S\ref{sec:outb}). 
In Section~\S\ref{sec:results} we present the timing and spectral analysis that resulted in the discovery of coherent pulsations and characterization of its broadband X-ray spectrum from \nustar and \nicer data. Finally from monitoring \textit{Fermi}/GBM and \nicer data obtained within the first 20 days of the outburst we can put strong constraints on the NS magnetic field strength (see \S \ref{sec:discussion}).

\section{The 2019 outburst of RX J0209.6-7427}
\label{sec:outb}

On 2019 November 20, the MAXI/GSC nova alert system triggered on an uncatalogued X-ray transient source \citep{2019ATel13300....1N}. Follow-up observations with the \textit{Neil Gehrels \swift Observatory} were performed on 2019 November 21, providing 
a localization of the system \citep{2019ATel13303....1K} at 
$\alpha_\text{J2000}=02^\text{h}09^\text{m} 33\fs{}85$ and
$\delta_\text{J2000}=-74^\circ 27' 12\farcs5$, with a 2.8\arcsec\ positional uncertainty. This position lies 2.9\arcsec from the known HMXB \rxj.
On 2019 November 21, \nicer began observations of this target that continue till the submission of this paper \citep{2019ATel13309....1I}. During the same period the outburst was detected by \textit{Fermi}/GBM, while \nustar made a single 
Directors Discretionary Time (DDT) observation.

\subsection{Data analysis}
\label{sec:data}

Below we provide basic information for the tools and methodology used for data extraction and analysis of the X-ray data obtained with \nustar, \nicer and \textit{Fermi}/GBM during the 2019 outburst of \rxj.

\subsubsection{NuSTAR}
The Nuclear Spectroscopic Telescope Array (\nustar) mission is the first focusing high-energy X-ray telescope in orbit operating in the band from 3 to 79 keV \citep{2013ApJ...770..103H}.
\nustar observed the system on 2019 November 26 with a 22~ks  DDT observation (obsid: 90502352002, MJD start:58813.33760417).
\nustar data were analysed with version 1.8.0 of the \nustar data analysis software (DAS), and instrumental calibration files from {\tt CalDB} v20191008. The data were calibrated using the standard settings on the {\tt NUPIPELINE} script, reducing internal high-energy background, and screening for passages through  the South Atlantic Anomaly \citep[see similar procedure in][]{2019A&A...621A.118K}. Using the {\tt NUPRODUCTS} script we extracted phase-averaged spectra for source and background regions (60\arcsec\ radius), as well as instrumental responses for each of the two focal plane modules (FPMA/B). Finally, for timing studies we performed barycentric corrections to event time of arrivals.

\subsubsection{NICER}
The \nicer X-ray Timing Instrument \citep[XTI,][]{2012SPIE.8443E..13G,2016SPIE.9905E..1HG} is a non-imaging, soft X-ray telescope aboard the \textit{International Space Station}. The XTI consists of an array of 56 co-aligned concentrator optics, each associated with a silicon drift detector \citep{2012SPIE.8453E..18P}, operating in the 0.2--12 keV band. The XTI provides high time resolution ($\sim100$  ns) and spectral  resolution of $\sim 85$ eV at  1 keV. It has a field of view of $\sim 30$ arcmin$^2$ in the sky and effective area of $\sim$1900 cm$^2$ at 1.5 keV (with the 52 currently active detectors). 

For the current study we analysed \nicer data obtained between MJD 58808--58834.
Data were reduced using {\tt HEASOFT} version 6.26.1, \nicer DAS version 2019-06-19\_V006a, and the calibration database (CALDB) version 20190516. 
For the timing analysis, we selected good time intervals according to the following conditions: 
\textit{ISS} not in the South Atlantic Anomaly region, source elevation $>20^\circ$ above the Earth limb ($>30^\circ$ above the bright Earth), pointing offset $\le 54$ arcsec, and magnetic cutoff rigidity (\texttt{COR\_SAX}) > 1.5 GeV/c.
For timing analysis, we performed barycentric corrections to event time of arrivals using the \texttt{barycorr} tool and the JPL DE405 planetary ephemeris.

For spectroscopy, we extracted spectra obtained quasi-simultaneously with \nustar (i.e., MJD 58813.382--58813.707, total exposure 865 s), in order to perform a broadband spectral fit. 
We generated the background spectrum from a grid of \nicer blank-sky spectra corresponding to the blank-sky pointings of Rossi X-ray Timing Explorer \citep[see][]{2006ApJS..163..401J}. This grid of spectra is populated with observed spectra in various space-weather observing conditions (K. C. Gendreau et al., in preparation). The background spectrum is generated by combining these blank-sky spectra weighted according to space-weather conditions and magnetic cutoff rigidities common to both the pulsar and background-fields observations.
For spectral fitting we used the latest available redistribution matrix and ancillary response files (v1.02).

\subsubsection{Fermi GBM}

The Gamma-ray Burst Monitor (GBM) on board \textit{Fermi} \citep{2009ApJ...702..791M} is an all sky monitor consisting of 12 sodium iodide (NaI) detectors and two bismuth germanate (BGO) detectors. 
The NaI detectors are sensitive to hard X-rays from 8--1000 keV while the BGOs extend this energy to 40~MeV and are not sensitive to typical accreting binaries.  There are three public data types available: CTIME which has 0.256 second timing resolution and 8 energy channels and is typically used for localization, transient detection, and this pulsar work, CSPEC which has 8 second timing resolution and 128 energy channels and is typically used for spectroscopy, and CTTE which is time tagged event data in 128 energy channels with 2 $\mu$s timing accuracy.

\section{Results}
\label{sec:results}

\begin{figure}
\includegraphics[width=1.0\columnwidth]{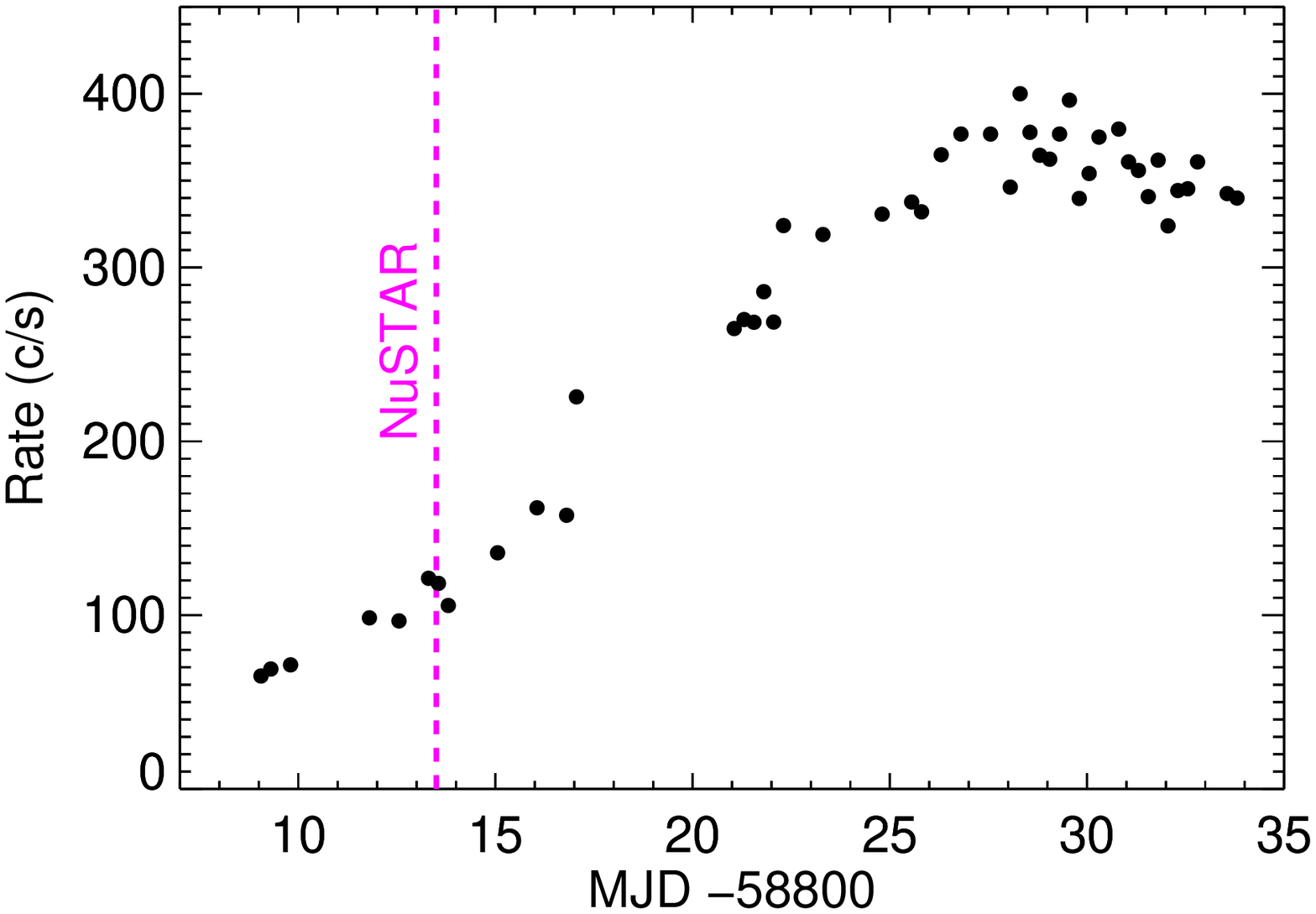}
\vspace{-0.5cm}
    \caption{X-ray lightcurve of \rxj based on \nicer count rates (0.5-8.0 keV) averaged over 6 hour intervals. The dashed vertical line indicates the epoch of the \nustar observation. First \nicer visit was performed within 1 day from the initial MAXI detection \citep{2019ATel13300....1N}. To translate \nicer count rates to bolometric $L_{\rm X}$ we find a conversion factor of $\sim4.75\times10^{36} \rm (erg/s)/(c/s)$ based on spectral fitting of the simultaneous obtained X-ray spectra (see \S~\ref{sec:spec} for details). 
    }
    \label{fig:LC}
\end{figure}

In Figure~\ref{fig:LC} we plot the X-ray lightcurve of \rxj during its 2019 outburst as obtained by \nicer. Given the brightness of the outburst no background subtraction was performed for the created lighcurve as its contribution is minimal ($<1\%$).
The outburst reached a peak flux in the \nicer band around 20 days after its initial detection.
For the current work we focus on the system's properties during a $\sim$25 day period after its original detection, which covers the rise of the outburst and is sufficient for our scientific goals. Further analysis of the data collected during the complete outburst will be presented in future publications.

\subsection{Spectral properties}
\label{sec:spec}

The X-ray spectra of BeXRBs show a power-law like shape having an exponential high energy cutoff (e.g., {\tt XSPEC} models `cutoffpl', `highecut', `bknpow', `fdcut') that originates from the accretion column \citep[e.g.,][]{2017A&A...605A..39T,2018MNRAS.480L.136M}. In many cases BeXRB spectra show residuals at soft energies. These residuals are often referred to as a ``soft-excess'' whose physical origin is attributed to a combination of mechanisms like, emission from the accretion disk, emission from the NS surface, or hot plasma around the magnetosphere  \citep{2004ApJ...614..881H}.
Due to the low foreground absorption, this ``soft-excess'' is often apparent in BeXRBs in the MCs \citep{2013MNRAS.436.2054B,2014MNRAS.444.3571S,2013A&A...558A..74V,2016MNRAS.461.1875V}.

\begin{figure}
\vspace{-0.3cm}
\resizebox{\hsize}{!}{\includegraphics*[angle=-90,trim=0 0 -80 0]{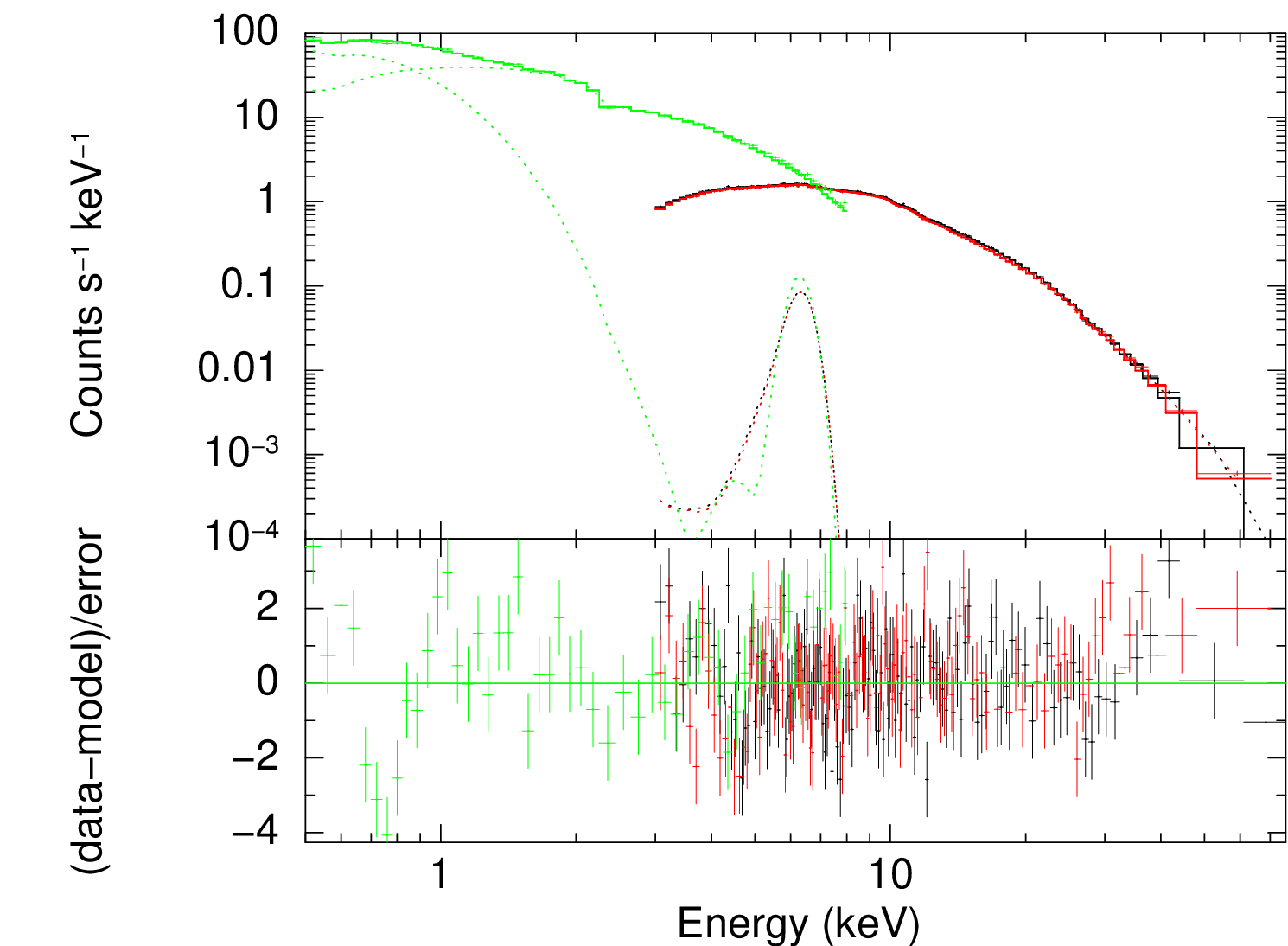}}
\vspace{-0.2cm}
    \caption{Phase averaged spectrum of \rxj (upper panel) during the \nicer/\nustar simultaneous observation. \nicer (green points) and \nustar (black and red points for FPM A and B) data are plotted together with the best fit model (see Table \ref{tab:spectral}). The individual spectral components (i.e. {\tt diskbb}, {\tt fdcut} and Gaussian line at 6.4 keV) are ploted with dashed lines .The lower panel shows the residuals to the best fit model. Events were re-binned for plotting purposes only.
    }
    \label{fig:nicer_nustar_spec}
\end{figure}

\begin{table}
\caption{Best-fit parameters of the empirical model\label{tab:spectral}}
\begin{threeparttable}[b]
\begin{tabular*}{\columnwidth}[t]{p{0.18\columnwidth}p{0.2\columnwidth}p{0.2\columnwidth}p{0.4\columnwidth}}
\hline
\hline\noalign{\smallskip} 
\multicolumn{4}{c}{{\tt xspec} model: {\tt Tbabs*(diskbb + fdcut + gaussian)}} \\
\hline
\hline
 & Parameter              &  Value   & Units  \\
\hline  
\hline\noalign{\smallskip}  
 & $N_{\rm{H}}$ Gal$^{(a)}$  & $1.58$ (fixed)  & $10^{21}$cm$^{-2}$\\
\\
{\tt diskbb} & k${\rm T_{BB}}$                  &$0.192_{-0.005}^{+0.007} $ & keV\\\noalign{\smallskip} 
 & {${\rm Norm_{BB}}$ $^{(b)}$}              &$5100_{-800}^{+700}$ &  $\rm (R_{BB}/D_{10})^2\cos{\theta}$ \\\noalign{\smallskip} 
 & {${\rm R_{BB}}$ $^{(b)}$}             &$470_{-40}^{+30}$ & km \\\noalign{\smallskip} 
\\
{\tt fdcut} & ${\rm \Gamma}$                      &${0.802}{\pm}{0.013}$  & - \\\noalign{\smallskip} 
 & ${\rm E_{\rm c}}$                  &$10.9_{-0.8}^{+0.9}$  & keV\\\noalign{\smallskip} 
 & ${\rm E_{\rm f}}$                 &$10.21_{-0.14}^{+0.16}$ & keV\\\noalign{\smallskip} 
 & ${\rm Norm}$  & $4.40_{-0.08}^{+0.06}$ & ${\times}10^{-2}$\\
\\
{\tt Gaussian} & $E_{\rm Fe}$                    &$6.34_{-0.08}^{+0.06}$  & keV\\\noalign{\smallskip} 
 & $\sigma_{\rm Fe}$ (keV)               &$0.33_{-0.06}^{+0.11}$  & keV\\\noalign{\smallskip} 
 & ${\rm Norm}$             &$3.4_{-0.4}^{+0.8}$   & ($10^{-4}$\,cm$^{-2}$\,s$^{-1}$)\\
\\
\multicolumn{4}{l}{Other information}\\
 & C$_{\rm FPMB}$ $^{(c)}$                        &$1.039_{-0.004}^{+0.003}$  & -\\\noalign{\smallskip} 
 & C$_{\rm NICER}$ $^{(c)}$                        &$0.989_{-0.010}^{+0.011}$  & -\\
\\
 & red.~$\chi^2 / {\rm dof}$             &$1.06906/2332$  & \\
\hline\noalign{\smallskip}
 & $\rm L_{\rm X}$  $^{(d)}$ &$5.54{\pm}{0.05}$  & \oergs{38}\\\noalign{\smallskip} 
 & $\dot{M}(L_{\rm X})$  $^{(e)}$     &$3.07{\pm}0.03$  & $10^{18}$g\,s$^{-1}$\\
  \hline\noalign{\smallskip}  
\end{tabular*}
\tnote{(a)} Galactic absorption was fixed to this value (see text for details).
\tnote{(b)} Disk radius was estimated from the normalization of the model, while assuming a disk inclination of 45$^\circ$ and distance of 55 kpc (i.e. $D_{10}=5.5$).
\tnote{(c)} The data from the 3 detectors were fitted simultaneously with all parameters tied apart from a constant that was left free to account for instrumental differences. 
\tnote{(b)} Unabsorbed X-ray luminosity in the (0.5--70 keV) band for a distance of 55 kpc.
\tnote{(d)} Mass accretion rate onto the NS, assuming $L_{\rm X}=0.2\dot{M}c^2$. 
\end{threeparttable}
\end{table}

We investigated the broadband spectrum of \rxj using simultaneous \nicer and \nustar data. Our goal is to fit the spectra using a phenomenological model and to measure the broadband (i.e., 0.5--70.0 keV) X-ray luminosity of the system. 
For spectral analysis all counts were regrouped to have at least 25 counts per bin.
Spectral analysis was performed using {\tt XSPEC} v12.10.1f \citep{1996ASPC..101...17A}.
Motivated by the spectral properties of BeXRBs, the \nicer and \nustar spectra were fitted simultaneously with a standard phenomenological continuum composed of a soft black body and a power-law with high energy cut-off. 
To account for the photoelectric absorption by the interstellar gas we used {\tt tbabs} in {\tt xspec} with Solar abundances set according to  \citet{2000ApJ...542..914W} and atomic cross sections from \citet{1996ApJ...465..487V}. 
Column density was fixed\footnote{We also tested a combination of two absorption components to account for Galactic absorption and intrinsic absorption as it is typical for BeXRBs in the Magellanic Clouds \citep[e.g.][]{2013A&A...558A..74V,2016MNRAS.461.1875V,2017MNRAS.470.1971V,2018MNRAS.475..220V}. 
The second component was left free to account for the absorption near the source or within the SMC, thus elemental abundances were fixed at 0.2 solar \citep{1992ApJ...384..508R}. However, the second component was not constrained by the fit and its column density was consistent zero, thus was not used in the reported spectral fit.} 
to the Galactic value of $1.58 \times10^{21}$ cm$^{-2}$ \citep{1990ARA&A..28..215D}.

Among the tested power-law models, the best fit was obtained by `fdcut' \citep{1986LNP...255..198T}. This model is a smoothed power-law with cut-off and is expressed analytically as:
\begin{equation}
dN/dE = E^{-\Gamma}\left(1+\exp{\frac{E-E_c}{E_f}}\right)^{-1}.   
\end{equation}
We note that non-smoothed models like `highecut' created sharp residuals around the cut-off energy that could be confused with absorption lines related to cyclotron scattering features that are often found in this energy range. A soft spectral component is needed to obtain an acceptable fit, we decided to use a typical disc black body ({\tt diskbb} in {\tt xspec}).
In the residuals of the fitted model there is clear evidence of an emission line present at $\sim6.4$  keV that originates from neutral Fe (K${\alpha}$ line).
The width of the Gaussian line ($\sim0.3$~keV) is comparable to that of other BeXRBs during major outbursts \citep[e.g., Swift\,J0243.6+6124;][]{2019ApJ...885...18J}. 
In regards to the best fitted model, there are still residuals around 1~keV that could be related to a mixture of emission lines as seen in other BeXRBs \citep[e.g., see SMC\,X-3,][]{2018A&A...614A..23K}. Investigating the nature of these residuals is beyond the scope of the paper\footnote{Due to limited calibration of \nicer around 1 keV it is quite possible that the origin of these features is a mixture of physical and instrumental effects.}. 
To estimate uncertainties we used Markov Chain Monte Carlo sampling method (available through {\tt xspec}). We used the Goodman-Weare algorithm to create a chain (total length of 10,000) of parameter values and create a probability distribution for each free parameter.
The best fit parameters (with their 90\% uncertainties) are presented in Table \ref{tab:spectral} and the spectrum is shown in Figure~\ref{fig:nicer_nustar_spec}. 
From the best fit model we are able to derive a conversion factor (i.e., $C_{\rm bol}$) to translate \nicer count rates (0.5--8.0 keV band) to ``bolometric'' X-ray luminosity (i.e., 0.5--70.0 keV) band. 
Assuming a distance\footnote{Although the distance to the SMC is found to be $\sim62$ kpc \citep{2014ApJ...780...59G}, for BeXRBs located at the SMC wing \citep[e.g.][]{2012MNRAS.420L..13H}, a distance of 55 kpc is commonly adopted \citep{2009AJ....137.3668C}.} to the source of 55 kpc \citep{2003MNRAS.339..157H}, we estimated $C_{\rm bol}\approx4.75\times10^{36} \rm (erg/s)/(c/s)$.

\begin{figure}
\vspace{-0.2cm}
\includegraphics[width=1.0\columnwidth]{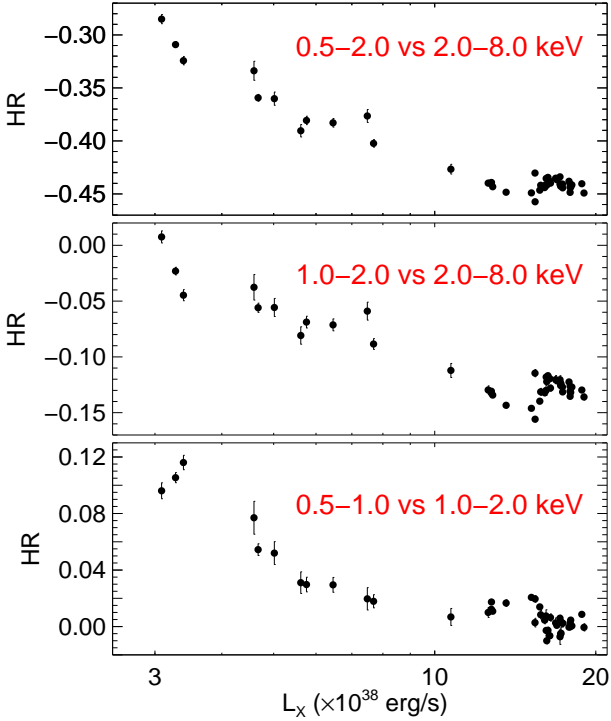}
\vspace{-0.9cm}
    \caption{Spectral hardness of \rxj as a function of X-ray luminosity. From top to bottom HRs are computed from \nicer data using different detector energy bands. For each point we summed events collected within a maximum of a 6 hours period.
    }
    \label{fig:LX_HR}
\end{figure}

To track the spectral evolution of \rxj during its 2019 outburst, we used the hardness ratio (HR), defined as $\rm{HR}=(\rm{R}_{i+1}-\rm{R}_{i})/(\rm{R}_{i+1}+\rm{R}_{i})$,   
where $\rm{R}_{i}$ is the count rate in a specific energy band. We split the collected events in 6 h intervals and computed HR using the 0.5--2.0 keV and 2.0--8.0 keV energy bands. 
We further computed HR indices using different \nicer energy bands. In the hard bands (1.0-2.0 vs 2.0-8.0 keV), HR is representative of changes of the power-law continuum. While in the soft bands (0.5-1.0 vs 1.0-2.0 keV), HR evolution is representative of the soft-excess and possible changes in absorption. We found no visual evidence of rapid changes in HR evolution that could be related with rapid increase in absorption (i.e. sudden increase in soft HR).
In Figure~\ref{fig:LX_HR} we plot HR as a function of the broadband $L_{\rm X}$ computed by translating \nicer count rates to $L_{\rm X}$ using $C_\mathrm{bol}$ \citep[see similar application to SMC X-3;][]{2017A&A...605A..39T}. Given that there is a spectral change with $L_{\rm X}$, this linear conversion from count rates to $L_{\rm X}$ should have an uncertainty of a factor of 30-50\%. Given that the spectrum becomes softer-when-brighter, around the peak of the outburst $L_{\rm X}$ should be overestimated by a small factor.

\subsection{Temporal properties}
\label{sec:temp}

In all  \nicer observations, a periodic modulation is apparent even by eye if we appropriately re-bin the events.
A pulsation search using PRESTO \citep{2002AJ....124.1788R} on the barycentered event data confirmed the coherent pulsations with a period of 9.29 s \citep[reported by][]{2019ATel13309....1I}. Pulsations near this period are detected in all \nicer observations as well as in the \nustar TOO observation. Following the period-based nomenclature introduced by \citet{2005MNRAS.356..502C} for BeXRB pulsars in the SMC, an alternative designation for \rxj is \sxp.

The pulse period has been decreasing during the outburst. We generated a phase-coherent timing model
for all the data analysed here (i.e. $\sim$25 days) using \textsc{Tempo2}  \citep{2006MNRAS.369..655H}. To do this,
we subdivided the data into intervals of less than 3000 seconds and generated one TOA per interval by cross-correlating a folded pulse profile with an
analytic template composed of 4 gaussians \citep{2011ApJS..194...17R}.
We fitted the TOAs to a timing model with two frequency derivatives, i.e. $\nu(t)=\nu_{\rm 0}+\dot{\nu}(t-t_{\mathrm 0})+(1/2)\ddot{\nu} (t-t_{\mathrm 0})^2$ while setting $t_{\mathrm 0}$ to 58810 MJD.
Nevertheless, there are substantial systematic residuals likely caused by torque noise.  We added a set of harmonically-related sinusoids (\texttt{WAVE} parameters in \textsc{Tempo2}) to obtain a model with nearly white
residuals. The results of the coherent timing analysis are summarized in Table \ref{tab:timing}.

\begin{table}
\caption{Pulse Timing Parameters for \rxj\label{tab:timing}}
\begin{tabular}{ll}
\hline\hline
\multicolumn{2}{c}{Fit and data-set} \\
\hline
MJD range\dotfill & 58808.9---58835.5 \\ 
Number of TOAs\dotfill & 85 \\
Rms timing residual (ms)\dotfill & 36 \\
\hline

\hline
\multicolumn{2}{c}{Set Quantities} \\ 
\hline
Right ascension, $\alpha$ (hh:mm:ss)\dotfill & 02:09:34.76 \\ 
Declination, $\delta$ (dd:mm:ss)\dotfill & $-$74:27:14.0 \\ 
Epoch of frequency determination $t_{\mathrm 0}$ (MJD)\dotfill &  58822 \\ 
\hline
\multicolumn{2}{c}{Measured Quantities} \\ 
\hline
Pulse frequency, $\nu_{\rm 0}$ (s$^{-1}$)\dotfill &  0.1075687(2) \\ 
First derivative of pulse frequency, $\dot{\nu}$ (s$^{-2}$)\dotfill &   1.165(3)$\times 10^{-10}$ \\ 
Second derivative of pulse frequency, $\ddot{\nu}$ (s$^{-3}$)\dotfill &  1.26(2)$\times 10^{-16}$ \\ 
\hline
\multicolumn{2}{c}{Whitening terms} \\ 
\hline
Reference epoch for Waves\dotfill &  58822 \\ 
Fundamental Wave frequency, $\omega_{\rm pw}$ (rad yr$^{-1}$)\dotfill &  48.0217 \\ 
Wave 1: $A_{\rm cos, 1}$; $A_{\rm sin, 1}$\dotfill & 24.6002; $-$25.2251 \\ 
Wave 2: $A_{\rm cos, 2}$; $A_{\rm sin, 2}$\dotfill & -21.6411; 42.6438 \\ 
Wave 3: $A_{\rm cos, 3}$; $A_{\rm sin, 3}$\dotfill & 19.1398; $-$28.4544 \\ 
Wave 4: $A_{\rm cos, 4}$; $A_{\rm sin, 4}$\dotfill & $-$4.15159; 11.3984 \\ 
Wave 5: $A_{\rm cos, 5}$; $A_{\rm sin, 5}$\dotfill & 0.183705; $-1.80281$ \\ 
\hline
\multicolumn{2}{c}{Assumptions} \\
\hline
Solar system ephemeris model\dotfill & DE405 \\
Time units \dotfill &  TDB \\
\hline
\end{tabular}
\end{table}

\begin{figure}
    \includegraphics[width=1.0\columnwidth]{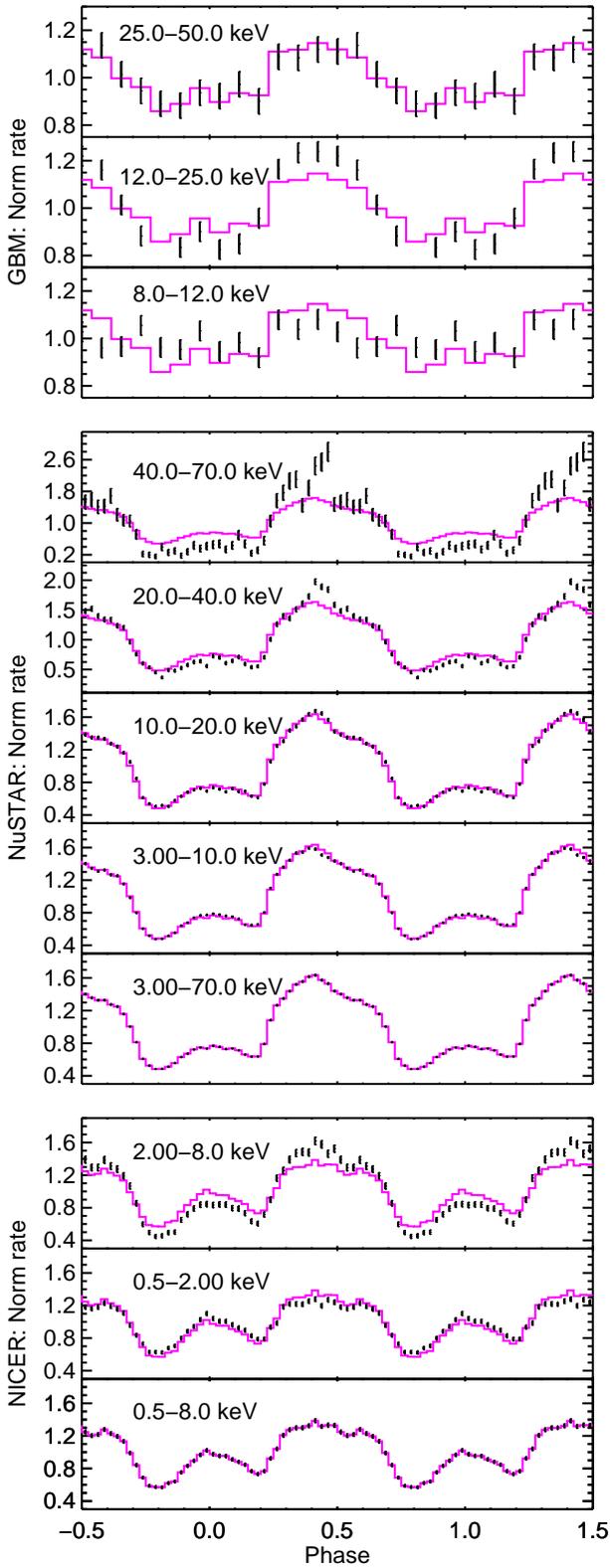}
    \vspace{-1.0cm}
    \caption{Energy resolved pulse profiles for \rxj obtained by quasi-simultaneous observations (see text). Each panel contains points from the corresponding energy band (see legend), while with magenta lines we plot the average pulse profile for each instrument. 
    }
    \label{fig:PPS}
\end{figure}

\begin{figure}
    \includegraphics[width=1.0\columnwidth]{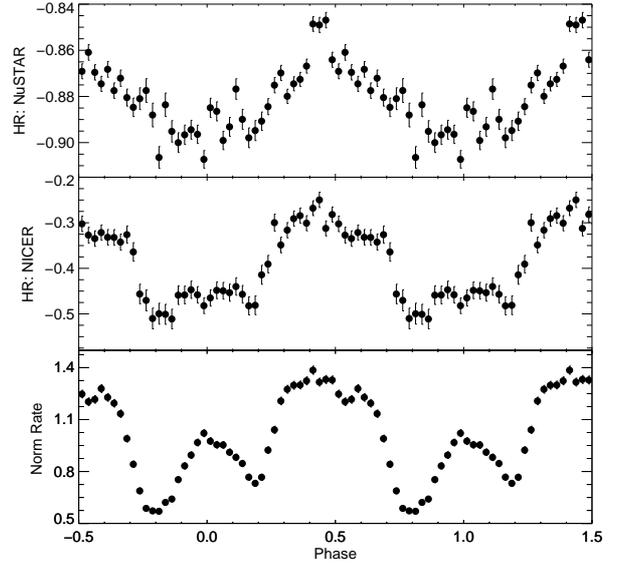}
    \vspace{-1.0cm}
    \caption{Pulse phase resolved hardness ratios of \rxj obtained by quasi-simultaneous observations (see text). For the \nustar HR (upper panel) we have used the 3.0--20.0 keV and 20.0--70.0 keV energy bands. For the \nicer HR (middle panel) we have used the 0.5--2.0 keV and 2.0--8.0 keV energy bands. In the lower panel we have plotted the average (0.5--8.0 keV) pulse profile obtained by \nicer. 
    }
    \label{fig:PP_HR}
\end{figure}

\begin{figure}
    \includegraphics[width=1.0\columnwidth]{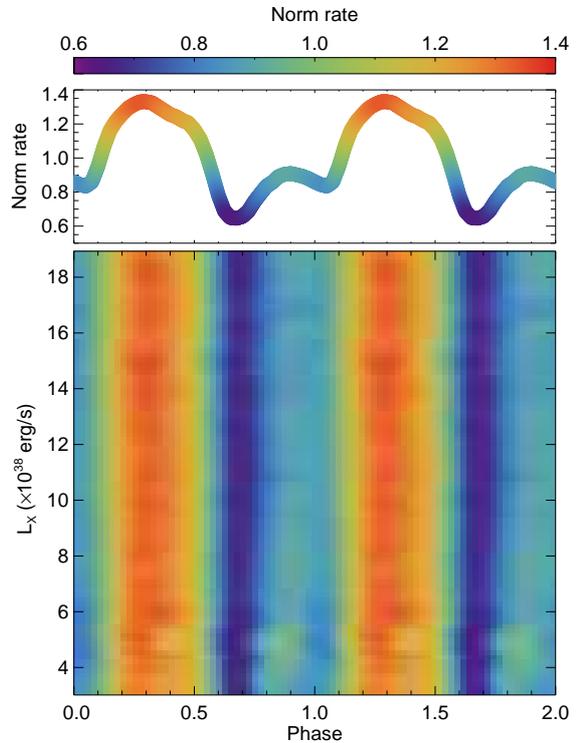} 
    \vspace{-1.0cm}
    \caption{\emph{Lower panel:} Heat map of the pulse profile evolution as a function of pulse-averaged $L_{\rm X}$ (0.5--70 keV) for the 25 first days of the outburst. The normalization is based on the $L_{\rm X}$ derived from the broadband spectral fit. \emph{Upper panel:} For comparison we have plotted the averaged pulse profile from the same period.
    }
    \label{fig:heat2}
\end{figure}

A search for pulsations was performed in the direction of \rxj in GBM data around the \nicer frequency in the manner described in \citet{1999ApJ...517..449F} and \citet{2012ApJ...759..124J}.
Specifically we used CTIME data products. Pulsation search was initially performed over 1000 frequencies from 106.91732 mHz to 108.07473 mHz as well as the next two harmonics in the energy range between 12.0-50.0 keV, over two day intervals. Once a phase model was established for the spin-up of the pulsar, the frequency search was repeated over the centroid frequency from the phase model plus and minus 50 increments of 925.9259$\times10^{-5}$ mHz. 
A search over 160 frequency derivatives was also added.  These where centered on the phase model frequency derivative at even intervals of 3.282$\times10^{-12}$ cycles/s/s.  The fitted Fourier amplitudes for the six harmonics with the frequency and frequency derivative that resulted in the maximum $Y_{n}$ statistic, as described in \citet{1999ApJ...517..449F}, and a $Y_{n}$ statistic that exceeded 35 were retained. 
The lowest $Y_{n}$ statistic was 59 for the first interval while the remainder of the detected frequencies had a $Y_{n}$ statistic exceeding 100 signifying a reliable detection.  A search with one day integrations was also performed and resulted in a poorer determination of the frequency derivative. The results are publicly available through the \fermi/GBM accreting pulsars project\footnote{See \url{https://gammaray.msfc.nasa.gov/gbm/science/pulsars/lightcurves/rxj0209.html}}.
To compare the temporal properties of \rxj obtained by \textit{Fermi}/GBM 
with those obtained by \nicer and \nustar we used GBM data within a one day period following MJD 58813-14.

To investigate changes of the pulse profile with energy we analysed all data obtained quasi-simultaneously with \nustar (see \S \ref{sec:data}). 
In Figure~\ref{fig:PPS} we present the pulse profiles for individual instrument bands, while in Figure~\ref{fig:PP_HR} we present the pulse phase resolved hardness ratios for \nicer and \nustar. Both \nustar and \nicer events were folded based on the same ephemeris.
The pulse profile is double peaked at low energies while it gradually changes to single peaked at high energies.

In order to visualize the pulse profile evolution we created a heat map of the pulse shape as a function of pulse averaged $L_{\rm X}$.
For this purpose we used only \nicer data. We split the events into 1 day intervals, assigned phases using the model in Table \ref{tab:timing} and  created pulse profiles using 40 phase bins. 
Each profile was smoothed and normalized with its average count rate. 
We then created a 2D histogram of the intensity of the system in pulse-phase and average pulse intensity. Finally, we converted \nicer count rates to broadband $L_{\rm X}$ by using the conversion factor obtained by the broadband spectral fit. The resulting heat-map is shown in Figure~\ref{fig:heat2}. The pulse-profile showed minimal evolution with $L_{\rm X}$, maintaining its double peaked shape. The only evident change was that the ``trough'' between the two main peaks (i.e. phase 1.0--1.1 in Fig. \ref{fig:heat2}) became shallower at large $L_{\rm X}$. The feature at 5\ergs{38} is due to lower statistics during that period. 

\subsection{NS magnetic field from spin-up}

Changes in the spin of a NS due to accretion can be predicted by theoretical models, if at least two parameters are known; the accretion rate (i.e., $\dot{M}$) and the surface magnetic field (i.e., $B$) of the NS \citep{1995ApJ...449L.153W}. Mass is transferred from the inner radius of a Keplerian disc that is truncated at the magnetospheric radius due to the balance of magnetic and gas pressures \citep{1977ApJ...217..578G}:
\begin{equation}
R_{\rm M} = \xi \left(\frac{R_{\rm NS}^{12}B^4}{2GM_{\rm NS}\dot{M}^2}\right)^{1/7},
\label{eq1}
\end{equation}
where $G$ is the gravitational constant, $M_{\rm NS}$ and $R_{\rm NS}$  is the NS mass and radius and $\xi\sim 0.5$ \citep{2018A&A...610A..46C} is a scaling factor between magnetospheric radius ($R_{\rm M}$) and Alfvén radius for disc accretion. 

As material is deposited onto the magnetic pole of the NS \citep[see][]{2007ApJ...654..435B},
the bolometric X-ray luminosity emitted can be converted to a mass accretion rate $\dot{M}$ assuming some efficiency $\eta_{\rm eff}$ (i.e., $L_{\rm X}\approx \eta_{\rm eff}\dot{M}c^2$).
This is generally assumed to be the efficiency under which gravitational energy is converted to radiation, namely $L_{\rm X}=GM_{\rm NS}\dot{M}/R$. For $R=R_{\rm NS}=10^6$~cm and $M_{\rm NS}=1.4M_{\odot}$, one finds $L_{\rm X}\approx0.2\dot{M}c^2$ (Henceforth, we adopt $\eta_{\rm eff}=0.2$).

The induced torque due to the mass accretion is 
$N_{\rm acc}\approx\dot{M}\sqrt{GM_{\rm NS}R_{\rm M}}$. 
The total torque can be expressed in the form of $N_{\rm tot}=n(\omega_\mathrm{fast})N_{\rm acc}$ where $n(\omega_\mathrm{fast})$ is a dimensionless function that accounts for the coupling of the magnetic field lines to the accretion disc  and takes the value $\approx7/6$ for slow rotators  \citep[for more details see][]{1995ApJ...449L.153W,2016ApJ...822...33P}.
The spin-up rate of the NS is then given by: 
\begin{equation}
\dot{v}=\frac{n(\omega_\mathrm{fast})}{{\rm 2 \pi} I_{\rm NS}} \dot{M} \sqrt{G M_{\rm NS} R_{\rm M}},
\label{eq2}
\end{equation}
where $I_{\rm NS} \simeq (1-1.7)\times10^{45}$~g cm$^{2}$ is the moment of inertia of the NS \citep[e.g.,][]{2015PhRvC..91a5804S}.
Henceforth, we adopt $I_{\rm NS} \simeq 1.3\times10^{45}$~g~cm$^{2}$.

\begin{figure}
    \includegraphics*[width=\columnwidth]{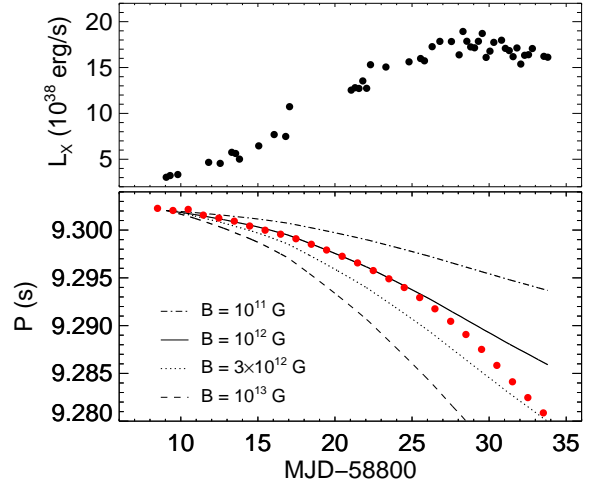}
    \vspace{-0.5cm}
    \caption{\emph{Upper panel:} X-ray light-curve of \sxp (0.5-70 keV), where $L_{\rm X}$ is estimated by scaling the  \nicer count rates with the correction factor estimated by the simultaneous \nicer and \nus spectral fit.
    \emph{Lower panel:} Spin period measurements based on \glast/GBM data (red points). With various lines we mark the predicted evolution of $P$ for different values of the NS magnetic field (see text for details).
    }
    \label{fig:P_evol_B}
\end{figure}

The spin period (or frequency) evolution of the NS can then be derived by solving Equation~\ref{eq2} for a variable $\dot{M}(t)$ and a constant $B$ value. The details of the methodology we followed are presented in the study of the spin-period evolution of NGC 300 ULX-1 \citep{2019MNRAS.488.5225V}.
In the case of \rxj the evolution of $\dot{M}$ during the outburst can be derived from the observed \nicer light curve assuming the scaling factor (count rate to $L_{\rm X}$) derived from the broadband spectral fit (see \S \ref{sec:spec}), and for various values of $B$. The derived spin-period evolutionary tracks are plotted together with the observed values in Figure~\ref{fig:P_evol_B}. 
From the figure it is clear that the observed evolution of $P$ is consistent with $B=$1--3$\times10^{12}$~G.  Given the fact that we used a linear conversion from \nicer count rates to $\dot{M}$ our estimation should have a systematic uncertainty and thus B could be underestimated by a factor of 2.
Moreover, it seems that the observed evolution of $P$ does not show any visual signature related to orbital Doppler shifts. As we will discuss in the next paragraph, although we cannot measure the orbital parameters, we can put constraints on the orbital period and perhaps inclination of the binary plane. 

\subsection{Constraints on binary orbital parameters from spin-up}

\begin{figure*}
	\includegraphics*[width=2\columnwidth]{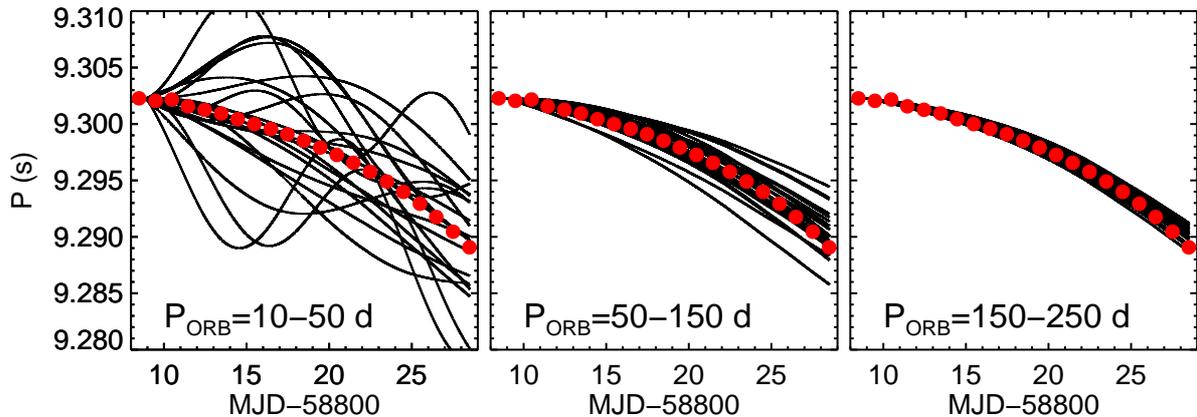}
	\vspace{-0.5cm}
    \caption{Fiducial evolution of observed period based on random orbital periods and inclination of the orbital plane in respect to the observer (see text). Similar to Figure~\ref{fig:P_evol_B}, we calculated the accretion torques based on observed $L_{\rm X}$ and $B=2\times10^{12}$ G. The lack of any apparent orbital signature in the data indicate that the orbital period of the system should be larger than $\sim$50 d, or that the binary system is observed face on.}
    \label{fig:P_evol_orb}
\end{figure*}

As of December 16 2019, the spin frequency of the NS in \rxj is increasing continually, while its evolution is consistent with spin-up due to accretion.
Specifically between MJD 58810-58826 the spin evolution of the NS is consistent with the predictions of spin-up due to accretion, while between  MJD 58826-58835 we see a deviation that is probably associated with orbital Doppler shifts.
Although an orbital signal is not yet evident in the data we can perform an exercise to demonstrate the effect of a fiducial binary orbit on the observed period of the NS. We simulated circular binary orbits and estimated the radial velocity of the NS, and thus the change in the observed period due to Doppler shifts. The mass of the NS was kept constant to 1.4$M_{\odot}$, while other parameters were assigned from uniform distributions; i.e., the mass of the donor star ($M_{\rm star}\in [8,10]~M_{\odot}$,) the orbital period ($P_{\rm orb}\in [10,250]~d$) and the binary plane inclination in respect to the observer ($\theta\in [0,90]^o$). In Figure~\ref{fig:P_evol_orb} we plot the observed periods from \glast/GBM together with predictions of random orbital models.

\section{Discussion}\label{sec:discussion}

We studied properties of the SMC BeXRB \rxj based on X-ray data collected during a $\sim$25 day period of its 2019 giant outburst.
Analysis of the \nicer and \glast/GBM data yielded the discovery of its spin period ($\sim9.3$~s). The pulse profile of the NS is double peaked at low energies while its pulsed fraction increases at higher energies (see Figure~\ref{fig:PPS}). During the evolution of the outburst there was minimal change in its pulse profile at low energies (see Figure~\ref{fig:heat2}). Broadband spectroscopy performed on \nicer and \nustar data enabled us to approximate its spectrum with a phenomenological model which provided a good estimate of a correction factor to transform \nicer count rates (0.5--8 keV) to broadband $L_{\rm X}$ (0.5--70 keV). 
We used \glast/GBM data to follow the spin evolution of the NS and compare it with the theoretically predicted spin-up due to accretion. We thus concluded that the surface magnetic field of the NS should be $\sim$1--3$\times10^{12}$ G, while its orbital period is most probably larger than 50 days.  

The spectral shape of accreting X-ray pulsars depends on the accretion regime.
The dependence of the spectral hardness with $L_{\rm X}$ has been shown to follow two different regimes \citep[e.g.,][]{2013A&A...551A...1R}. In the subcritical regime the spectrum of the pulsar becomes harder when brighter, while in the supercritical regime the opposite behavior takes place.
For the pulsars studied by \citet{2013A&A...551A...1R}, the critical $L_{\rm X}$ where the turn from subcritical to supercritical regime occurs around $1-2$\ergs{37} (0.3--10.0 keV band). The transition should be related to a formation of a shock above the NS hot spot, where material is deposited \citep{2007ApJ...654..435B}. While for different BeXRBs, this transition should provide hints for the magnetic field strength of the NS, as the cross-section for the scattering of in-falling material is a function of $B$ \citep{1971PhRvD...3.2303C,1974ApJ...190..141L}.
However, it should be noted that the softer-when-brighter dependence only refers to the phase-average properties. The opposite behaviour occurs within the pulse phase, where the spectrum becomes harder with increased pulsed intensity (see Figure~\ref{fig:PP_HR}). This is quite typical to X-ray pulsars and is related to the anisotropic emission from the accretion column. 
In fact the double peaked pulse profile of \rxj is characteristic of BeXRBs in the supercritical regime, as is typically considered a signature of radiation escaping from the sides of the accretion column in a fan-beam emission pattern  \citep{1975A&A....42..311B}.

In Figure~\ref{fig:LX_HR} we show the spectral hardness as a function of $L_{\rm X}$. 
The source behavior at $L_{\rm X}=(2-10)$\ergs{38} is consistent with the supercritical regime, i.e., the diagonal branch of the hardness-intensity diagrams \citep{2013A&A...551A...1R}.
In addition, the figure indicates the presence of perhaps a third branch that appears above \oergs{39}, which is similar to the spectral evolution of SMC X-3 during its 2016 outburst \citep{2018A&A...614A..23K}. In that case, \citet{2018A&A...614A..23K} claimed that this stabilization might have eluded detection because the sources studied by \citet{2013A&A...551A...1R} never reached such high $L_{\rm X}$. It is plausible that the spectral hardness stabilization is a manifestation of physical changes in the accretion column; i.e, the accretion column reached its maximum height and/or the optical depth of the in-flowing material exceeded unity. 
Moreover, the HR does not have the necessary sensitivity to trace complicated changes in the spectral shape. 
For example, the ``soft-excess'' typically becomes brighter with $L_{\rm X}$ thus resulting in softer HR, a stabilization of the HR could be a result of the ``soft-excess'' reaching a saturation limit \citep[see SMC\,X-3 case,][]{2018A&A...614A..23K}. In addition the high energy cut-off of the spectrum typically moves to lower energies as the source optical depth in the accretion column becomes higher. 
A detailed study of these effects is beyond the scope of this paper.

The spectral (i.e., softer-when-brighter evolution) and temporal (i.e., double peaked pulse profile and spin-evolution) properties of \rxj are evidence of the system remaining in the super-critical regime during the observed period.
At this point we can estimate the critical $L_{\rm crit}$ where the accretion column is formed resulting in a pivot point in the spectral hardness evolution with $L_{\rm X}$.
Following \citet{2012A&A...544A.123B} this is given by:
\begin{equation}
L_{\rm crit}=\left(\frac{B}{0.688\times10^{12}~G}\right)^{16/15}{\times}10^{37}~{\rm erg/s},
\label{eq3}
\end{equation}
that holds for typical parameters for the NS mass ($M_{NS}=1.4M_{\odot}$), radius ($R_{NS}=10$~km), standard disk accretion, and an accretion column where the seed photons inside the column originate from bremsstrahlung emission \citep[see eq. 32 of][for more details]{2012A&A...544A.123B}. For $B=1-3\times$10$^{12}$~G, equation~\ref{eq3} yields a critical $L_{\rm crit}{\sim}1.5-4.4$\ergs{37}. 
Monitoring observations during the decay of the outburst could verify our estimated $B$ value through more detailed timing analysis (i.e., if an orbital modulation is found), or spectral transition in the context of \citet{2013A&A...551A...1R}. 

Another mechanism that is often used to probe the dipole $B$ strength of the NS in BeXRBs is the propeller transition \citep{1975A&A....39..185I}.
Although, the transition between accretor regimes and propeller is often missed due to observational sampling \citep[e.g.,][]{2017MNRAS.470.1971V},
we generally expect to observe a sharp drop in the observed flux when this occurs \citep[e.g.,][]{1996ApJ...457L..31C,1997ApJ...482L.163C,2016A&A...593A..16T}. Assuming $B=1-3\times10^{12}$~G for \rxj we found a limiting luminosity ($L_{\rm X,Lim}$) of $1-9$\ergs{35} before the onset of propeller transition \citep[see eq. 3 of][]{2018A&A...610A..46C}. 
For comparison for $B=10^{13}$ G we would expect $L_{\rm X,Lim}\sim$\oergs{37} (i.e., $L_{\rm X}\propto{B^2}$).

In the context of PULXs, \rxj is yet another example of a system that can reach super-Eddington luminosity even though the NS has a typical B field strength, which is in agreement with the properties of the majority of known PULXs \citep{2019MNRAS.485.3588K}. 
Moreover, there is no change in the pulse profile of the system with $L_{\rm X}$, thus there is no evidence of beaming of the pulsed component as the BeXRB luminosity exceeds super-Eddington limit.

\section{conclusion}

\rxj is a BeXRB system located in the outer SMC wing that exhibited a super-Eddington outburst in November 2019. The analysis of \nicer data revealed the presence of coherent pulsations with a period of $\sim9.3$ s. During the outburst we obtained simultaneous \nicer and \nustar observations that enabled us to perform broadband spectroscopy, thus characterizing its spectral shape and accurately measuring its X-ray luminosity. Moreover, no evidence of a cyclotron resonance feature was found in the \nustar spectrum of the source.
From \nicer monitoring data of the outburst we found that NS reached a peak luminosity of $\sim2$\ergs{39} (0.5--70 keV), momentarily making \rxj a ULX pulsar, and perhaps the brightest BeXRB ever observed in the SMC. 
Furthermore, we used \glast/GBM data to follow the spin evolution of the NS and compare it with the theoretically predicted spin-up due to accretion. We thus concluded that the surface magnetic field of the NS is $\sim$3$\times10^{12}$ G and its orbital period is likely $\gtrsim$50 days.

\section*{Acknowledgements}
The authors would like to thank the anonymous referee for the the constructive report that helped to improve the manuscript.
This work was supported by NASA through the \nicer mission and the 
Astrophysics Explorers Program.
This work was supported by NASA through the Fermi Guest Investigator Program. Facilities: \nicer, \nustar, \glast.
We acknowledge the use of public data from the \swift\ data archive.
A.L.S. is supported by an NSF Astronomy and Astrophysics Postdoctoral Fellowship under award AST-1801792.
D.A. acknowledges support from the Royal Society.




\bibliographystyle{mnras}
\bibliography{general}








\bsp	
\label{lastpage}
\end{document}